\documentclass{moriond}

\def\Journal#1#2#3#4{{#1} {\bf #2}, #3 (#4)}

\newcommand{\apj}{Astrophys. J.}   %
\newcommand{\mnras}{Mon. Not. R. Astron. Soc.}   %
\newcommand{\pasp}{Publ. Astron. Soc. Pac.}   %

\def\be{\begin{equation}}
\def\ee{\end{equation}}
\def\bea{\begin{eqnarray}}
\def\eea{\end{eqnarray}}

\newcommand{\Photo}{\includegraphics[height=35mm]{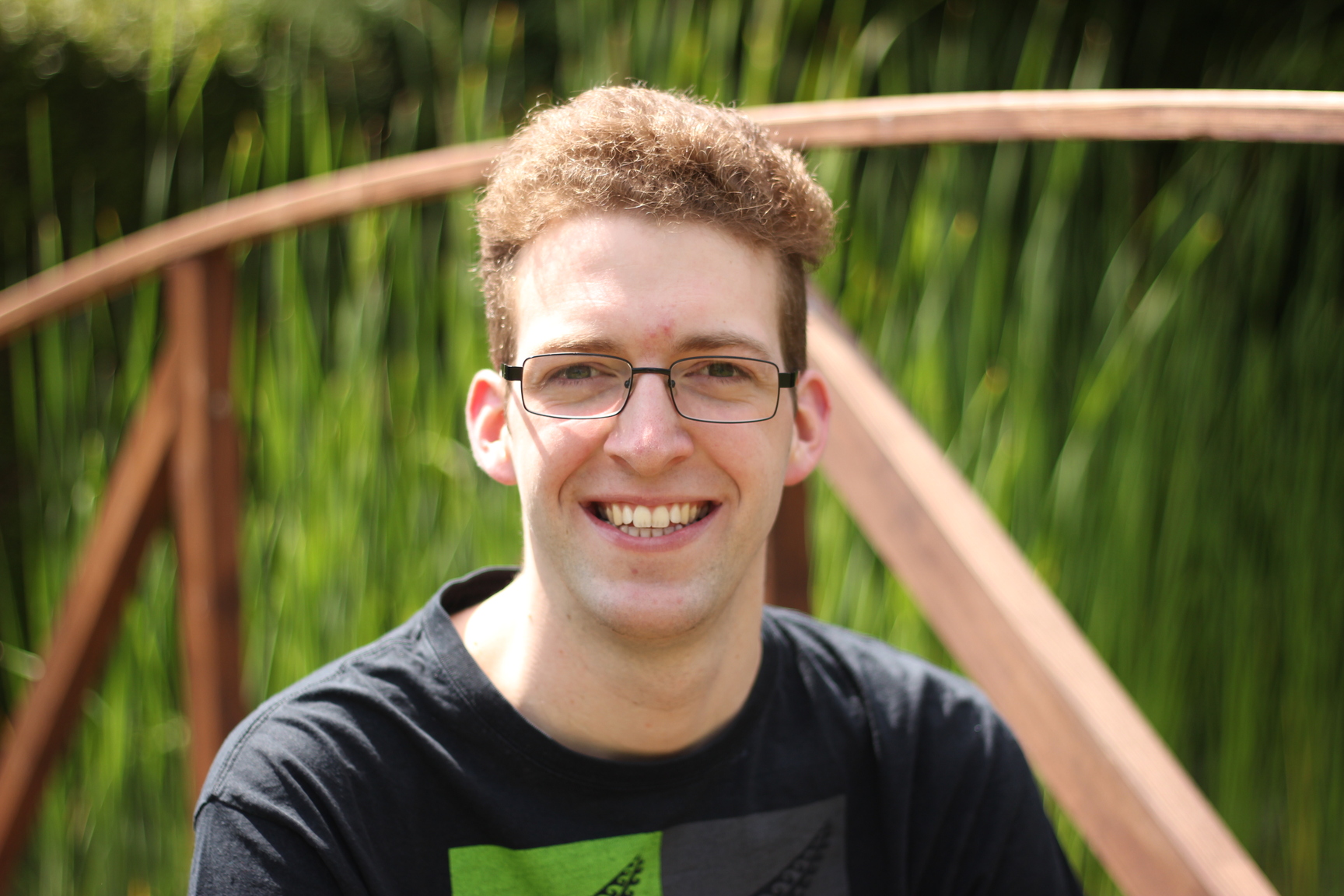}}
\usepackage{amssymb}
\usepackage{caption}
\usepackage{xcolor}

\begin{document}
\vspace*{4cm}
\title{\textcolor{gray}{Contribution to the 2022 Cosmology session of the 56th Rencontres de Moriond}\vspace{1em}\\Reionization constraints from HERA 21cm power spectrum limits}

\author{ Stefan Heimersheim}

\address{Institute of Astronomy, University of Cambridge,\\
Madingley Road, Cambridge CB3 0HA, UK}

\vspace{-2em}%
\maketitle\abstracts{
I present our analysis of the 21cm power spectrum upper limits from the HERA radio interferometer, published in HERA Collaboration et al. 2022 \cite{H21A}. We use the recent limits to constrain reionization and the properties of the IGM and galaxies in the early universe. I focus in particular on the possibility of a radio background in addition to the CMB (e.g. produced by early galaxies) which can lead to a stronger 21cm signal and is thus easier to constrain. I show what limits the HERA observations can put on these models and the IGM, and how this compares to existing constraints on the radio and X-ray background.
}

\section{Introduction}
Over the last decades we have explored both, the nearby Universe of quasars and galaxies, as well as the furthest edge of the observable Universe with the Cosmic Microwave Background. 21cm cosmology aims to fill the gap in between, probing
the neutral hydrogen from Cosmic Dawn to the Epoch of Reionization, and now it is a focus of many instruments.

The key observational target is the signature of the 21cm hydrogen line seen against the background radiation. The line
corresponds to the neutral-hydrogen hyperfine transition which can cause absorption or emission at the (redshifted) 1.42\,GHz frequency. This deviation from the smooth spectrum of the background radiation can, in principle, be observed today.
One of the telescopes aiming to detect the 21cm signal is the HERA radio interferometer in South Africa \cite{DeBoer}.

The strength of the 21cm signal is determined by the 21cm \textit{brightness temperature,} $T_{21}$, which depends on the
difference between the background radiation intensity (\textit{radiation temperature,} $T_{\rm rad}$) and the Hydrogen spin distribution (\textit{spin temperature,} $T_S$), illustrated in Figure \ref{fig0}. The latter couples to the gas temperature $T_K$ as soon as Ly-$\alpha$
photons are abundant (via the Wouthuysen-Field effect), which is usually the case for signals detectable by HERA.
$T_{21}$ is proportional to $\left(1-T_{\rm rad}/T_S\right)$, allowing for a positive signal (increase over the background by up to $\sim 30\,\mathrm{mK}$), or a negative signal (absorbing a fraction of the background).

\section{Methods}

\subsection{Data}
The results presented here are based on the first public HERA data
release \footnote{\url{https://reionization.org/science/public-data-release-1/}},
measurements taken with 39 science-quality antennas over 18 nights of observations.
In particular we use the data originating from a low-foreground part of the sky (HERA \textit{Field 1}).

A detailed description of the measurements and derived power spectra is provided in
the HERA Upper Limits paper \cite{H21L}, but the key takeaway from
the analysis is that the measurements constitute \textit{upper limits} on the value of the cosmological 21cm power spectrum
$\Delta^2_{21}$ as the observed values include non-cosmological ``systematics'' contributions.
Marginalizing over possible systematic contributions we obtain a likelihood $L_m$ for a given theoretical power spectrum $\Delta^2_{21}$
which is essentially a smoothed step function at the observed values.

\subsection{Analysis}
We model the 21cm power spectrum using a semi-numerical simulation most recently updated by Reis, Fialkov and Barkana \cite{Reis}.
The simulation covers a volume of 384 comoving Mpc side length with 3 comoving Mpc resolution and redshifts from Dark Ages
($z\sim60$) to $z=5$.
We vary a number of parameters such as the minimum circular velocity $V_c$ of star-forming halos and the fraction of collapsed gas forming stars $f_*$. Together these parameters determine the amount of stars and star forming halos, and thus scale the emission of radiation.
Next we select the X-ray efficiency $f_X$ and radio efficiency $f_r$ relative to the present day population, and finally vary the amount of ionizing radiation emitted, mapping to a certain reionization optical depth $\tau$.

To put constraints on these 5 parameters we need to vary all parameters at the same time, allowing us to derive marginalized 1D and 2D posteriors for every parameter and combination.
For this we use a Markov Chain Monte Carlo (MCMC) method, specifically the ensemble sampler
\texttt{emcee} \footnote{\url{https://emcee.readthedocs.io/}}. To analyze and plot the resulting
parameter samples we use the codes \texttt{anesthetic} \footnote{\url{https://anesthetic.readthedocs.io/}} and \texttt{GetDist} \footnote{\url{https://getdist.readthedocs.io/}}.
In this process we use a neural network (emulator) to predict the power spectrum from the parameters.
This essentially serves to interpolate between 10,000 existing simulation runs and significantly accelerates
our analysis. This emulation introduces a relative error of about 20\% but this is lower than the observational
uncertainty of our data and has little effect on the likelihood.
The full details of the simulation and analysis methods as well as all our results can be found in the HERA Astrophysical Constraints paper \cite{H21A}.

\section{Results}
\begin{figure}[h]
\centering
\begin{minipage}{.4\textwidth}
  \centering
  \vspace{0.8cm}
  \includegraphics[width=\linewidth]{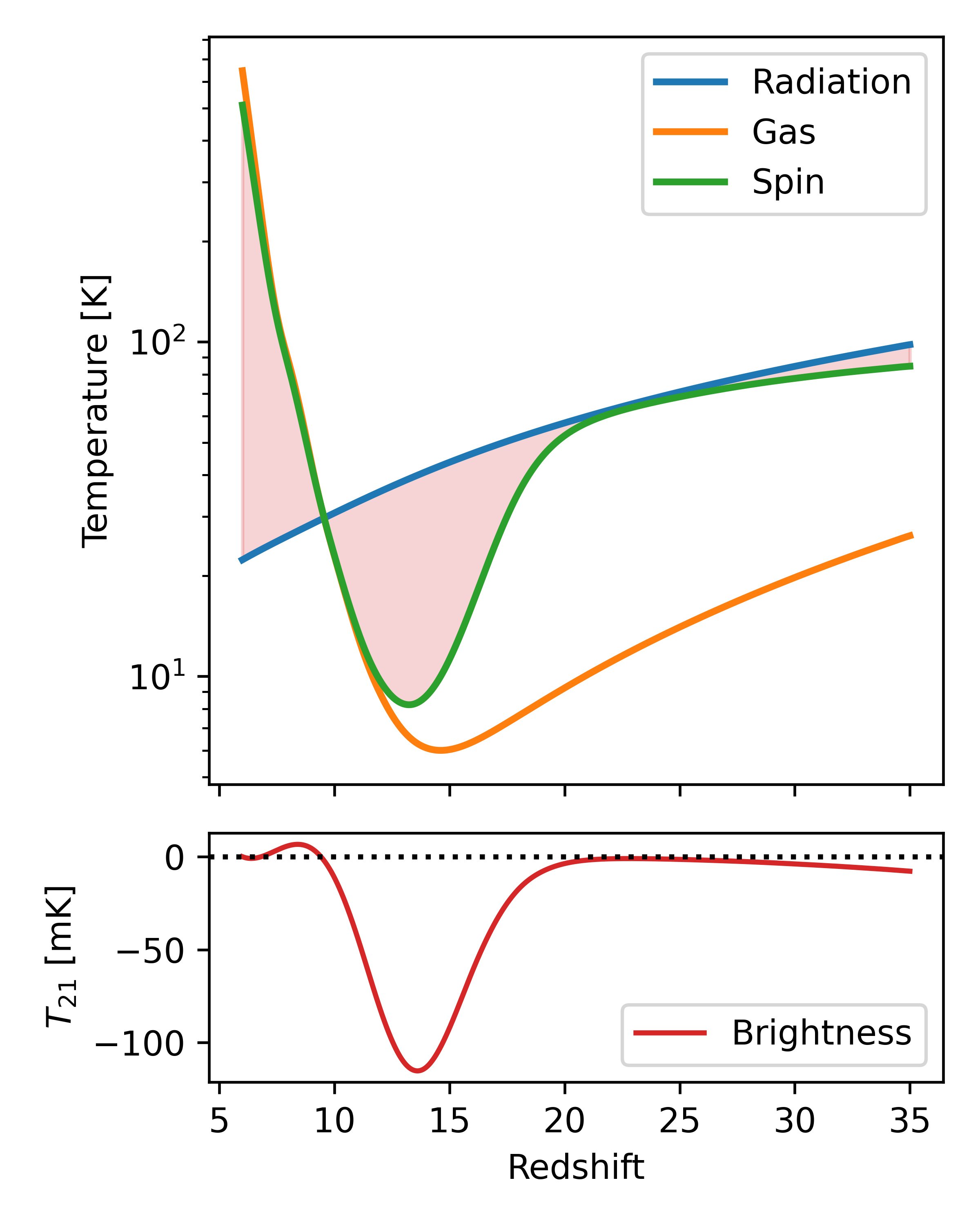}
  \captionof{figure}{Radiation (blue), gas (orange) and spin (green) temperature evolution with redshift, illustrating the contrast (red area) that determines the 21cm brightness temperature (red line below).}
  \label{fig0}
\end{minipage}
\hfill
\begin{minipage}{.58\textwidth}
  \centering
  \includegraphics[width=\linewidth]{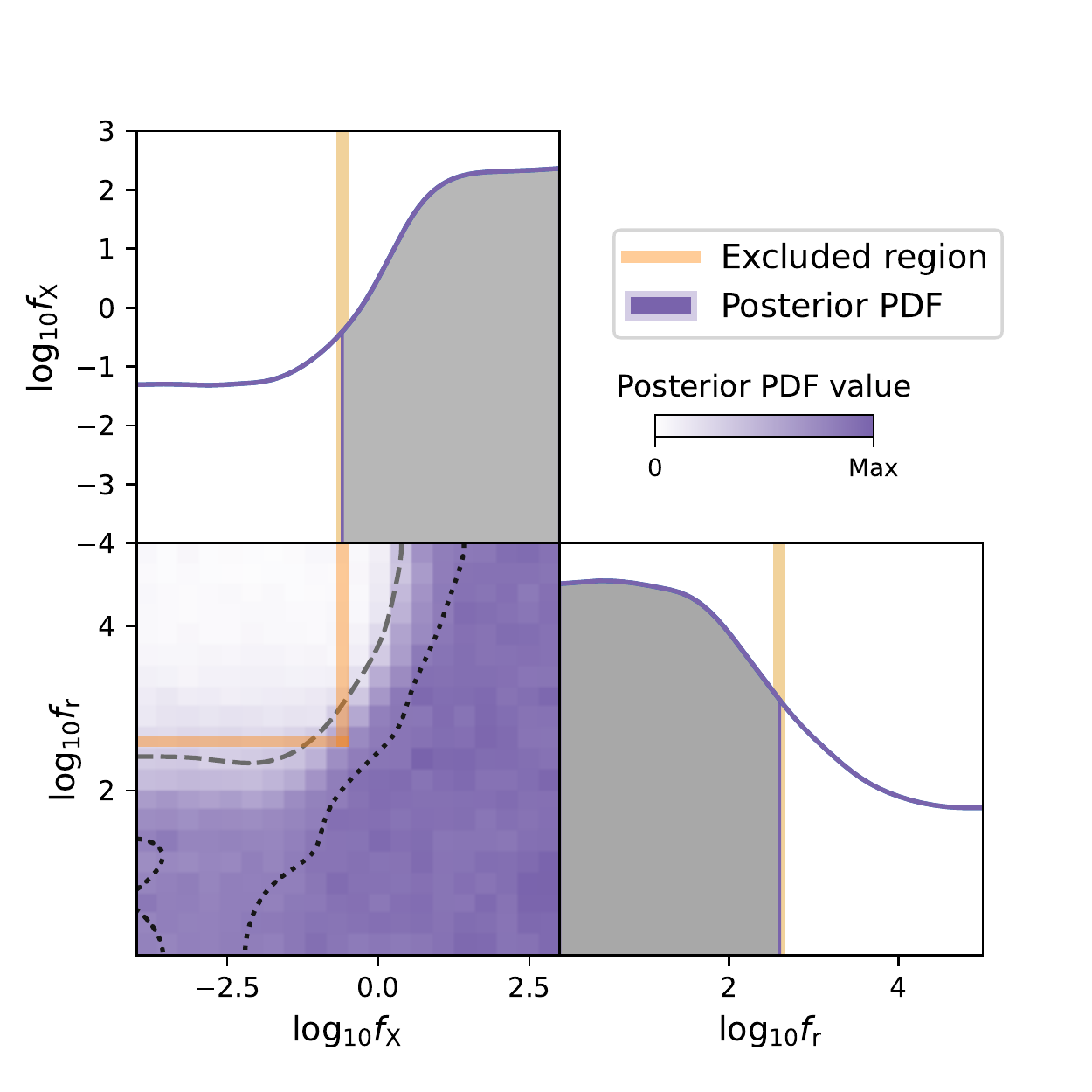}
  \captionof{figure}{1D and 2D marginalized posterior probability distributions for the two most relevant model parameters, $f_X$ and $f_r$. The purple colour indicates the probability density, while the black lines show 68\% and 95\% 2D iso-probability contours.}
  \label{fig1}
\end{minipage}
\end{figure}

We focus here on models which include the option for extra radio background sourced by early galaxies, and show what
constraints the current HERA limits can provide on these models.
Out of the 5 parameters we vary, only two have a significant effect on the posterior. The X-ray heating efficiency $f_X$
and the radio efficiency $f_r$ have large effects on the 21cm power spectrum as they determine the contrast between
the gas temperature (heated by X-rays) and the background radiation (boosted by radio galaxies), and thus how large
the 21cm signal and power spectrum can be (see Figure \ref{fig0}).

Figure \ref{fig1} shows the posterior distribution of these two parameters, as well as their 2D posterior histogram.
The 2D histogram clearly shows that most models with $f_X<0.25$ and $f_r>397$ (orange lines, approximately
corresponding to the 95\% confidence contours) are ruled out independently of the other parameters. This is corroborated by the 1D 
posteriors, indicating that values of $f_X<0.25$ or $f_r>397$ are disfavoured at 68\% confidence level.

\begin{figure}[h]
    \centering
    \includegraphics[height=2.6in]{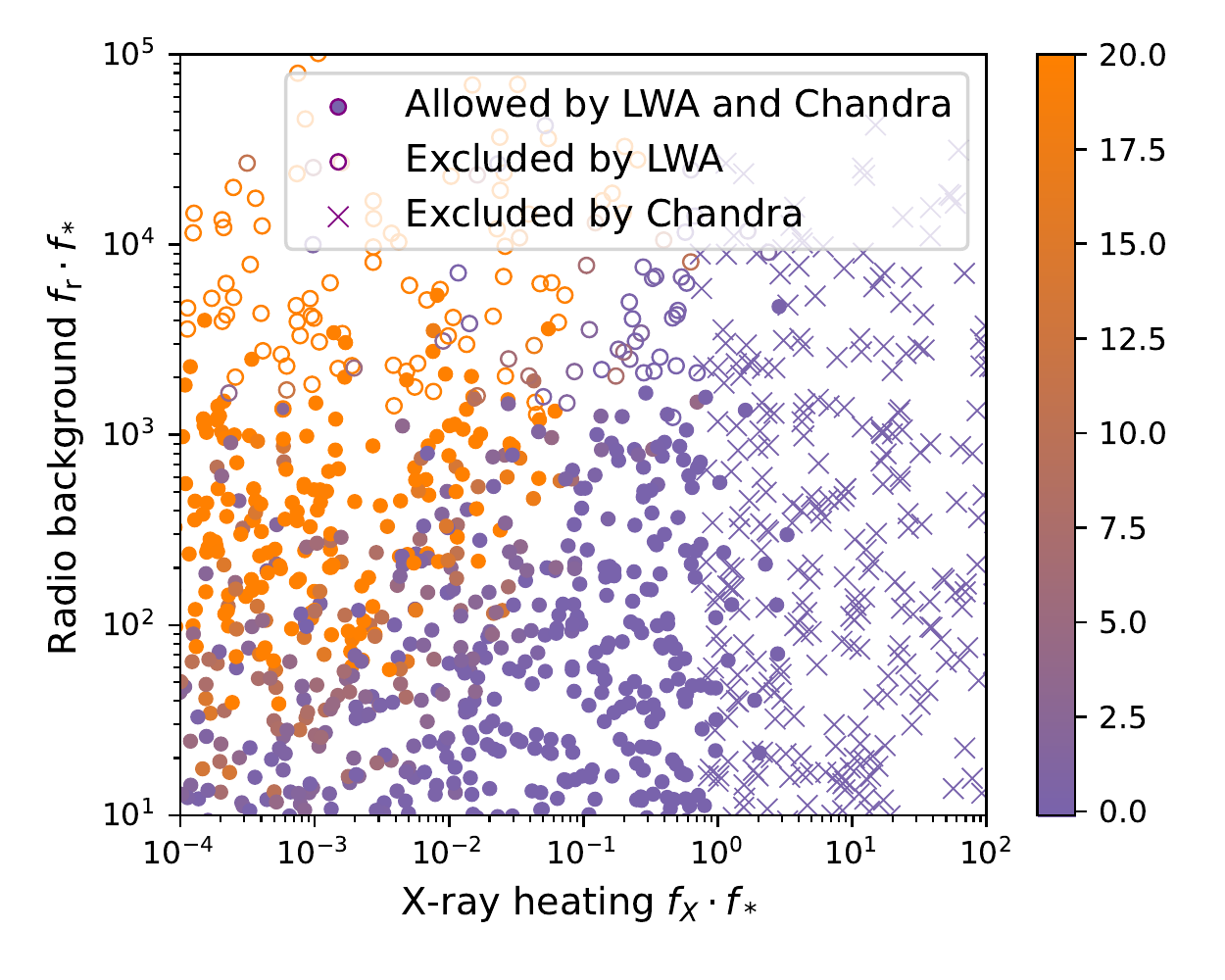}
    \hfill
    \includegraphics[height=2.6in]{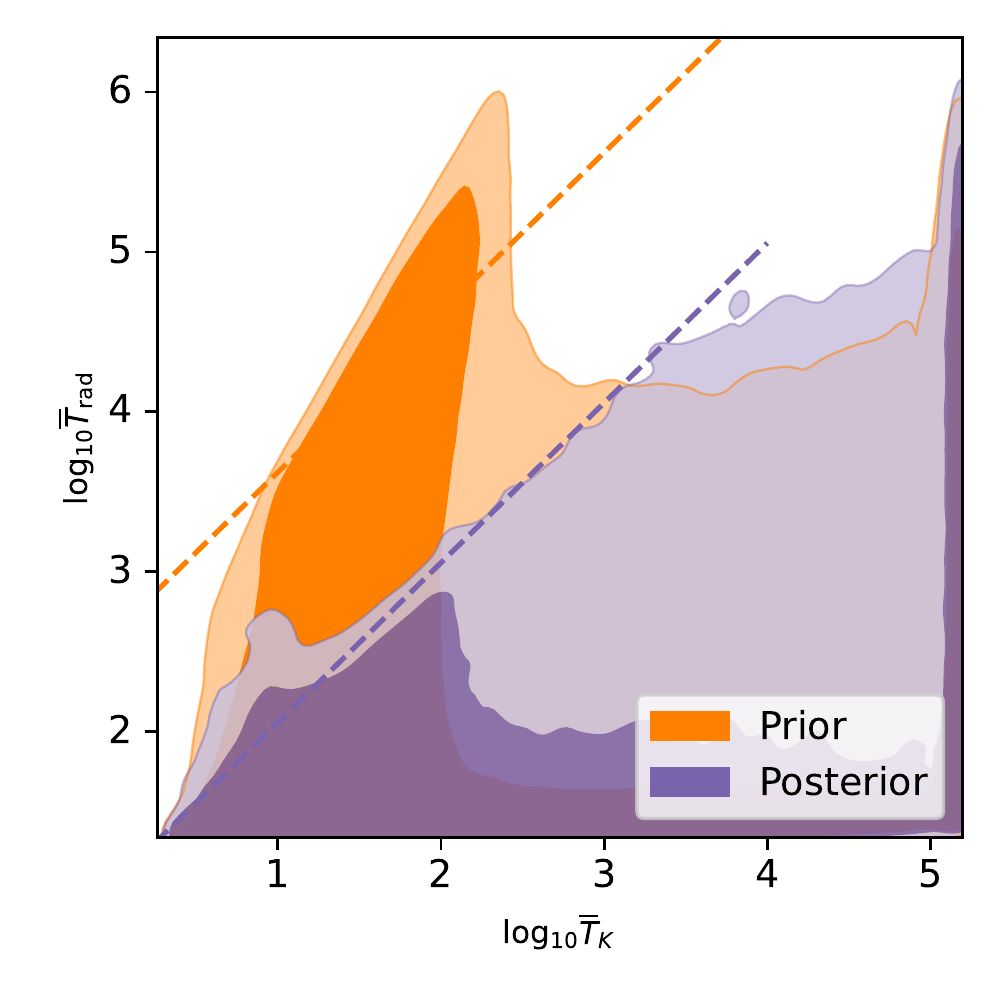}
    \caption{\textbf{Left:} Model parameter samples in $f_r\cdot f_*$--$f_X\cdot f_*$ space illustrating the LWA/ARCADE2 and Chandra constraints on parameter space. The colour indicates the HERA likelihood, showing models ruled out at $\gtrsim 4\sigma$ in orange.
    \textbf{Right:} Prior and posterior distribution of modelled radiation and gas temperatures at $z=8$.}
    \label{fig2}
\end{figure}

To put these values into context we consider the level of radio and X-ray background observed today. The ARCADE2 experiment
gives a limit on the radio background in excess of the CMB (approximately scales with $f_*\cdot f_r$), and the Chandra observations of unresolved X-ray background impose a bound on the amount of X-ray background produced (roughly proportional to $f_*\cdot f_X$). We impose these bounds as conservative constraints on our models, checking whether the models produce backgrounds exceeding these limits by redshift 8.
Figure \ref{fig2} (left) contains samples in this parameter space, showing a fraction of models excluded as they exceed the Chandra limit (marked with crosses), some models allowed by Chandra but exceeding ARCADE2/LWA measurements (empty circles), and the remaining parameter points as filled circles.
We demonstrate which models can be constrained by HERA using the color scale, points in purple are compatible with HERA while points in orange are ruled out with a loglikelihood difference $\Delta\log L_m>20$.
HERA can clearly constrain a significant fraction of the models allowed by Chandra and ARCADE2/LWA, and complement these experiments.

In Figure \ref{fig2} (right) we switch from model parameters to IGM properties and deliver a constraint on temperatures of the intergalactic gas $\bar T_K$ and the background radiation $\bar T_{\rm rad}$. Unlike in the case of model parameters, our priors here are not uniform so we show the prior distribution of temperatures in our model (orange) as well as the posterior (purple) after applying the HERA constraints. We see a clear bound on the ratio between radiation- and gas temperature; large ratios of $\log_{10}(\bar T_{\rm rad}/\bar T_K)>1.1$ are ruled out at 95\% confidence level.

Finally we briefly discuss standard models, without any radio background beyond the CMB,
constrained by HERA together with complementary probes.
In this case the strongest signals can be achieved with a cold IGM (little heating) and thus HERA
mainly provides a lower limit on the the X-ray luminosity per star formation rate,
constraining $L_{X<2\,\mathrm{keV}}/SFR$ to be in $[10^{40.2}, 10^{41.9}]$ at 68\% confidence level.
More details on this as well as further constraints can be found in the HERA paper \cite{H21A}.

\section*{Acknowledgments}

My presentation is based on work done together with the HERA Collaboration, summarized in a collaboration paper \cite{H21A}. The work I focused on here would not have been possible without the support of the whole collaboration, especially my supervisor Anastasia Fialkov, and Steven R. Furlanetto who led the HERA Theory team. The additional results presented in my presentation
are based on sections led by
Jordan Mirocha,
Julian B. Muñoz,
Nicholas S. Kern,
Steven G. Murray,
and Yuxiang Qin.
Beyond these headline names I want to explicitly thank everyone else who has contributed to the paper \footnote{Zara Abdurashidova,
James E. Aguirre,
Paul Alexander,
Zaki S. Ali,
Yanga Balfour,
Rennan Barkana,
Adam P. Beardsley,
Gianni Bernardi,
Tashalee S. Billings,
Judd D. Bowman,
Richard F. Bradley,
Philip Bull,
Jacob Burba,
Steve Carey,
Chris L. Carilli,
Carina Cheng,
David R. DeBoer,
Matt Dexter,
Eloy de Lera Acedo,
Joshua S. Dillon,
John Ely,
Aaron Ewall-Wice,
Nicolas Fagnoni,
Anastasia Fialkov,
Randall Fritz,
Steven R. Furlanetto,
Kingsley Gale-Sides,
Brian Glendenning,
Deepthi Gorthi,
Bradley Greig,
Jasper Grobbelaar,
Ziyaad Halday,
Bryna J. Hazelton,
Stefan Heimersheim,
Jacqueline N. Hewitt,
Jack Hickish,
Daniel C. Jacobs,
Austin Julius,
Joshua Kerrigan,
Piyanat Kittiwisit,
Saul A. Kohn,
Matthew Kolopanis,
Adam Lanman,
Paul La Plante,
Telalo Lekalake,
David Lewis,
Adrian Liu,
Yin-Zhe Ma,
David MacMahon,
Lourence Malan,
Cresshim Malgas,
Matthys Maree,
Zachary E. Martinot,
Eunice Matsetela,
Andrei Mesinger,
Mathakane Molewa,
Miguel F. Morales,
Tshegofalang Mosiane,
Abraham R. Neben,
Bojan Nikolic,
Chuneeta D. Nunhokee,
Aaron R. Parsons,
Nipanjana Patra,
Samantha Pieterse,
Jonathan C. Pober,
Nima Razavi-Ghods,
Itamar Reis,
Jon Ringuette,
James Robnett,
Kathryn Rosie,
Mario G. Santos,
Sudipta Sikder,
Peter Sims,
Craig Smith,
Angelo Syce,
and Nithyanandan Thyagarajan}.

\end{document}